\documentclass[a4paper,11pt]{article}
\usepackage{pos}
\usepackage{subfig}

\def\kt{k_{\perp}}

\def\mutpr{\mu_\perp^{\prime}}
\def\mutp2{\mu_\perp^{\prime 2}}

\begin{document}

\title{Branching probabilities in TMD Monte Carlo algorithms}
\ShortTitle{Branching probabilities in TMD Monte Carlo algorithms}

\author{L.~Keersmaekers}

\affiliation{Elementary Particle Physics, University of Antwerp, Antwerp, Belgium}

\emailAdd{Lissa.Keersmaekers@uantwerpen.be}

\abstract{We discuss a recently proposed  
branching algorithm which incorporates transverse momentum 
dependent (TMD) parton splitting probabilities, and can be used 
for Monte Carlo event generators based on TMD distributions.} 

\FullConference{%
  41st International Conference on High Energy physics - ICHEP2022\\
  6-13 July, 2022\\
  Bologna, Italy
}

\maketitle

The parton branching formulation~\cite{Hautmann:2017xtx} 
for transverse momentum dependent (TMD) 
distributions~\cite{Abdulov:2021ivr,Angeles-Martinez:2015sea}  
has allowed one to explore the impact of TMD evolution 
on a variety of collider processes (for instance, 
deep inelastic scattering (DIS)~\cite{BermudezMartinez:2018fsv}, Drell-Yan (DY) 
lepton pair  production~\cite{BermudezMartinez:2019anj}, 
di-jets~\cite{Abdulhamid:2021xtt}, 
DY + jets~\cite{Yang:2022qgk,Martinez:2021chk}),  
  and  investigate the  role of   transverse momentum 
  recoils~\cite{Dooling:2012uw,Hautmann:2013fla}   and 
  soft-gluon angular 
  ordering~\cite{Marchesini:1987cf,Catani:1990rr,Hautmann:2019biw}  
   in Monte Carlo  parton showers. 
   
 In the above studies, the branching probabilities are 
given by splitting functions computed in the collinear approximation~\cite{dglapref1,dglapref2,dglapref3}.  
It has long been known, however, that in 
regions of phase space sensitive to infrared 
phenomena~\cite{Bassetto:1983mvz,Collins:2003fm}  
contributions beyond the collinear approximation 
can be relevant. 
  In this article, based on the work~\cite{Hautmann:2022xuc}, 
 we discuss    TMD splitting functions   defined from 
  high-energy factorization~\cite{Catani:1994sq}. These splitting functions take into account 
   finite transverse momentum tails in the branching probabilities, which become important 
   when the gluons exchanged in the initial-state (spacelike) parton decay chain   
carry small   longitudinal momentum fractions $x$, and may be treated as soft.   
In the following we describe the results of 
constructing a Monte Carlo branching 
algorithm~\cite{Hautmann:2022xuc,Keersmaekers:2021arn} 
which relies  on the implementation of TMD  splitting 
functions in the framework~\cite{Hautmann:2017xtx}. 

The starting point is the branching kinematics for the space-like 
parton shower~\cite{Gieseke:2003rz},  
including angular ordering of soft emissions.  
We describe the parton splitting process at  each vertex 
through the off-shell  TMD splitting 
functions computed in 
Refs.~\cite{Catani:1994sq,Hautmann:2012sh,Gituliar:2015agu,Hentschinski:2016wya,Hentschinski:2017ayz}, based on high-energy 
factorization~\cite{Catani:1990eg}.  
These splitting functions  are positive definite, and 
interpolate consistently between the 
collinear limit~\cite{dglapref1,dglapref2,dglapref3} and  
the high-energy limit~\cite{Kuraev:1977fs,Balitsky:1978ic}. 

In Ref.~\cite{Hautmann:2022xuc} we construct  
corresponding   TMD  Sudakov form factors, which are 
obtained from 
 the angular average $ {\overline P}_{ba}$ 
 of the  TMD splitting functions as 
\begin{equation} 
\Delta_a(\mu^2,\kt^2)=\exp\left(-\sum_b\int_{\mu_0^2}^{\mu^2}\frac{d\mu^{\prime 2}}{\mu'^2}\int_0^{z_M}dz\ z {\overline P}^R_{ba}(z,\kt^2,\mu^{\prime 2})\right), 
\label{eq:ap:sud}
\end{equation}
where $a, b$ are flavor indices, $\mu$ is the evolution scale, 
$z$ is the longitudinal momentum fraction, 
$\kt$ is the transverse momentum, 
and $z_M$ is the soft-gluon resolution scale, possibly 
dependent on $\mu$~\cite{Hautmann:2019biw} 
according to the angular ordering. 
Using the unitarity picture as in~\cite{Hautmann:2017xtx}, 
and requiring four-momentum conservation  in the parton 
decay chain, we obtain  branching equations for the 
TMD parton distributions   
$\tilde{\mathcal{A}}_a$~\cite{Angeles-Martinez:2015sea} 
of the form 
\begin{align}
\label{eveq2}
 \tilde{\mathcal{A}}_a\left( x,\kt^2, \mu^2\right) &= 
 \Delta_a\left(\mu^2,\kt^2\right)\tilde{\mathcal{A}}_a\left( x,\kt^2, \mu_0^2\right) + 
 \\
& \hspace{-1.9cm} 
 \sum_b\int\frac{d^2\mutpr}{\pi\mutp2}
 %
  \int_x^1 \textrm{d}z\,  
K_{ a b } ( z,  \mutpr, \kt , \mu , z_M ) 
  \tilde{\mathcal{A}}_b\left( \frac{x}{z},  (\kt+(1-z)\mutpr)^2, \mutp2\right).
  \nonumber 
\end{align}
where the kernel $K_{ a b }$ contains  the TMD splitting functions, 
Sudakov form factors and phase space constraints. 

 With the TMD splittings and form factors, in  Eq.~(\ref{eveq2}) 
 we aim   at a combined treatment of  small-$x$ and Sudakov 
contributions to parton evolution.  
This is 
relevant to describe the exclusive structure of 
jet final states  at high energies, see 
for instance~\cite{Dooling:2014kia,Hautmann:2008vd,Deak:2009xt,Deak:2011}. 
Other approaches 
to the treatment of Sudakov and small-$x$ effects have recently 
been investigated, see e.g. the 
study~\cite{Taels:2022tza}. 
The distinctive feature of the approach in 
Ref.~\cite{Hautmann:2022xuc} is that it works 
at the level of unintegrated, $k_\perp$-dependent splitting functions 
which factorize 
in the high-energy limit and control the summation of 
small-$x$ logarithmic contributions to 
the evolution. These  splitting functions are then used in the 
branching algorithm, 
where they are integrated to construct the new Sudakov factors. 
As a result, the TMD distributions 
fulfill  integral relations expressing 
the momentum sum rules. 

\begin{figure}[h!]
\begin{center} 
    \subfloat{
    \includegraphics[width=0.4\textwidth]{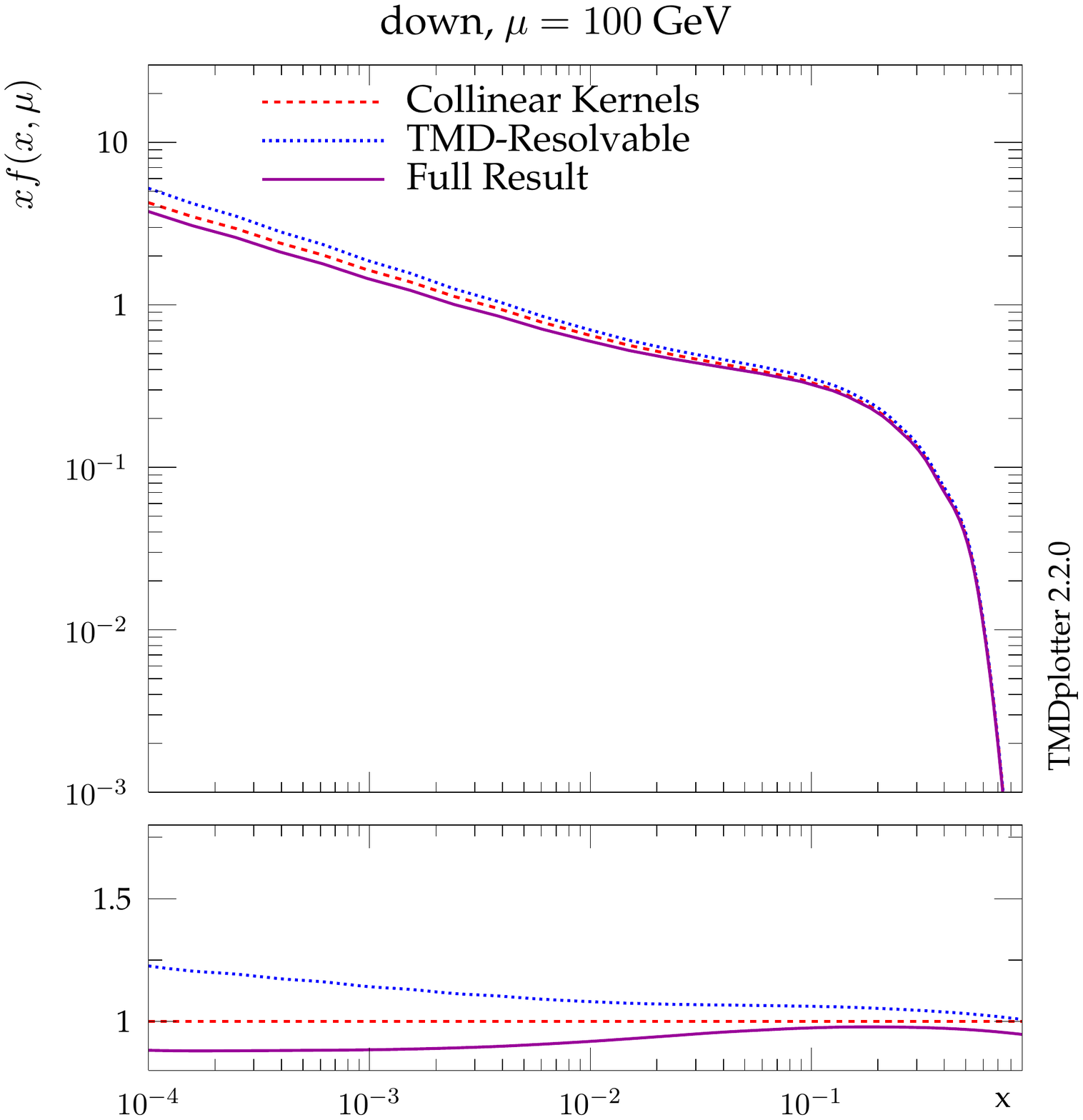}
    }
    \qquad
    \subfloat{
    \includegraphics[width=0.4\textwidth]{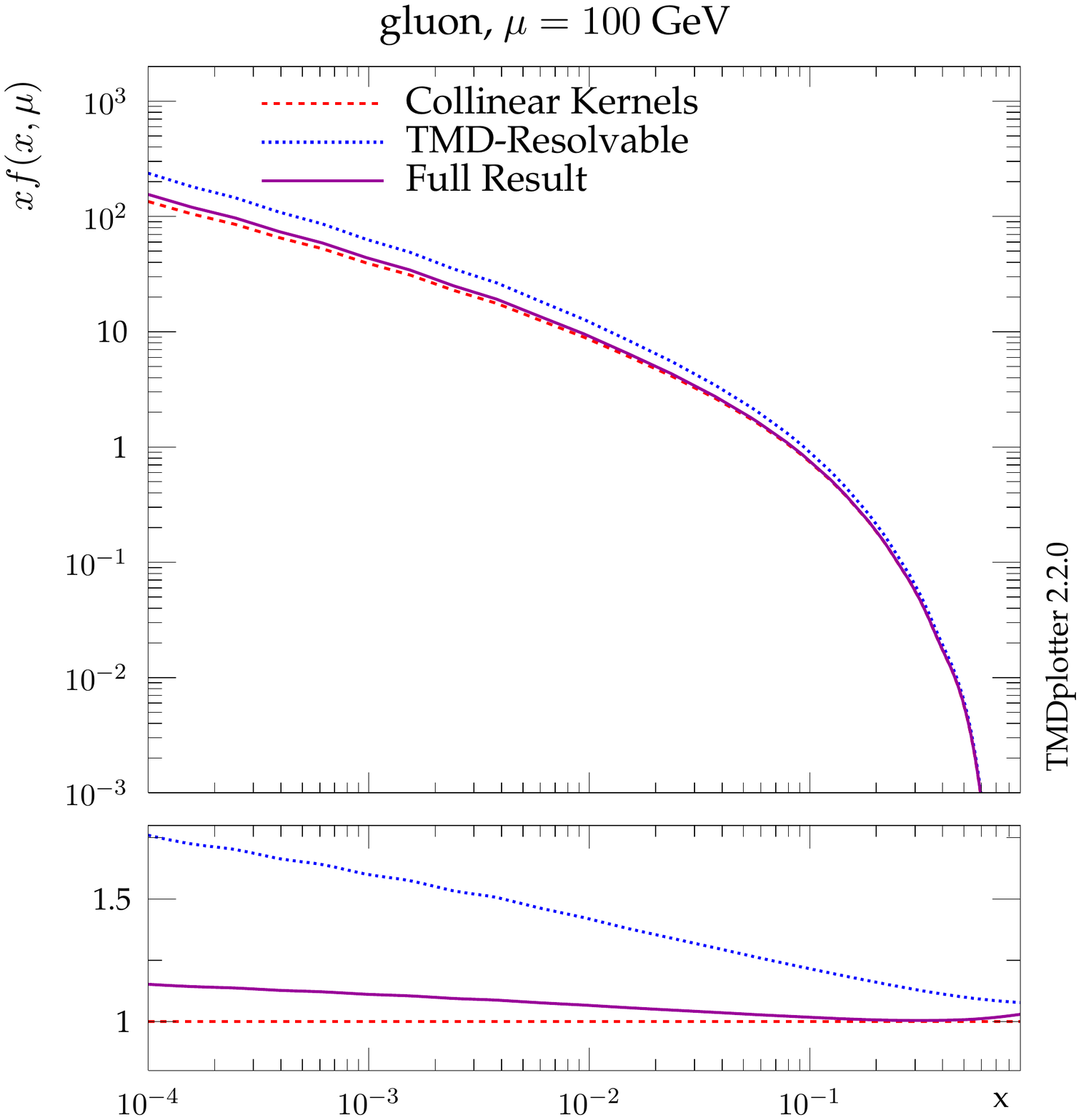}
    }
    \qquad
        \subfloat{
    \includegraphics[width=0.4\textwidth]{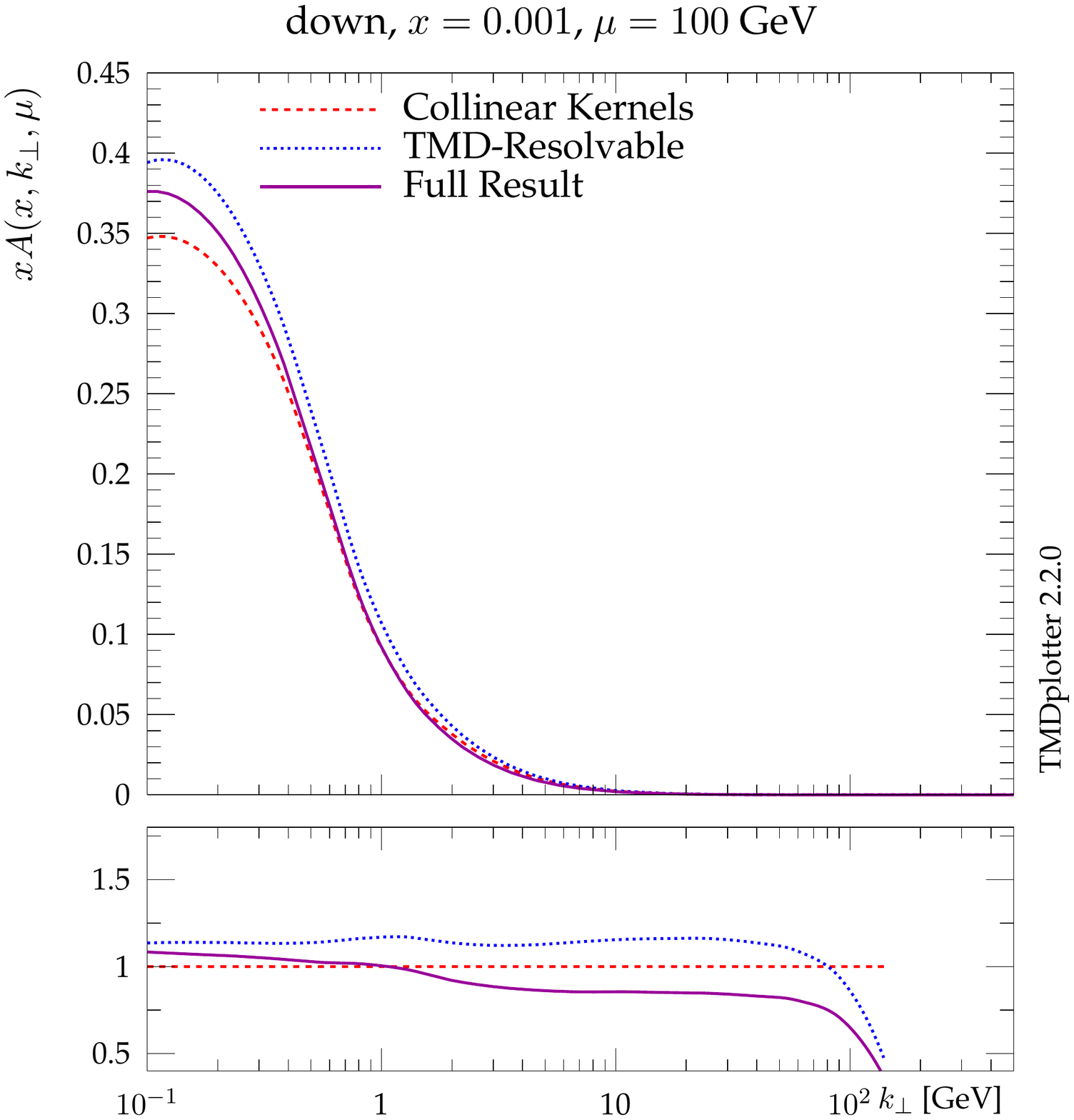}
    }
    \qquad
        \subfloat{
    \includegraphics[width=0.4\textwidth]{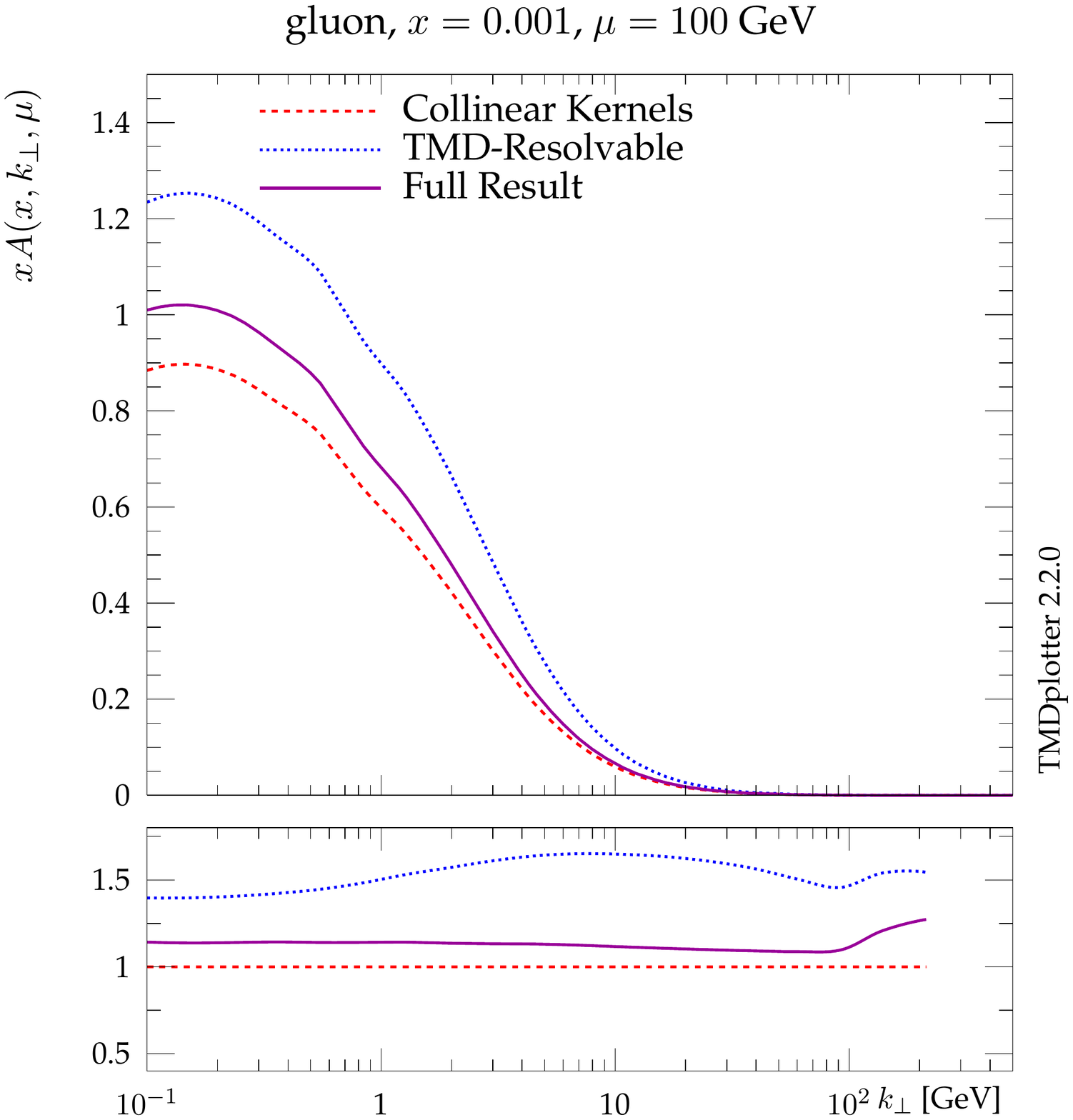}
    }
    \qquad
  \caption{\it    TMD (bottom) and integrated (top) 
   distributions  evolved by 
  using the new branching equations with 
  $k_\perp$-dependent splitting kernels 
  (solid magenta curves), compared with the    
  contribution  from  $k_\perp$-dependent splittings in 
  resolvable emissions only (dotted blue curves)  and  the result 
  of evolution with purely collinear splitting kernels (dashed red curves).}
\label{fig2:evolveddistributions}
\end{center}
\end{figure}

To illustrate the  effects of the branching evolution, 
in Fig.~\ref{fig2:evolveddistributions} we solve Eq.~(\ref{eveq2})  
by numerical Monte Carlo techniques for given boundary 
conditions, which we take to be the 
TMD parameterizations~\cite{BermudezMartinez:2018fsv} 
at  $\mu_0 = 1.4$ GeV. 
(Other parameterizations available e.g.~in the 
library~\cite{Abdulov:2021ivr} could also be used 
for the purpose of this illustration.)  
 The solid magenta curves in the top panels of 
 Fig.~\ref{fig2:evolveddistributions} 
 show the $x$ dependence of the 
 gluon and down-quark TMD 
distributions evolved to $\mu= 100 $ GeV 
 and integrated over $k_{\perp}^2$,
  while the solid magenta curves in the bottom panels  
show the $k_\perp$ dependence 
of the  gluon and down-quark  
distributions at $\mu= 100 $ GeV
for a fixed value of $x$. 
For comparison, in Fig.~\ref{fig2:evolveddistributions}  
we also plot 
 the 
results which are obtained with the same 
distributions at scale $\mu_0$ but 
without including any  $k_\perp$ dependence  in the splitting kernels, that is, with 
the purely collinear splitting kernels (dashed red curves), and  
the results which are obtained by including 
the $k_\perp$ dependence of splitting functions in 
resolvable emissions only (dotted blue curves). In contrast to the full result and the result with purely collinear kernels, the model with the $k_\perp$ dependent splitting functions in 
resolvable emissions only does not satisfy  momentum sum rules, which leads to a significant departure from the full result.
We see that the influence of the TMD splitting 
kernels on evolution is significant especially for  low $x$, gives rise to 
 a change in the  $k_\perp$  and $x$ shapes of the 
 distributions and does not disappear completely after integration over $k_\perp$. 

The numerical implementation of an approach 
which both includes the TMD splitting functions and 
satisfies the momentum sum rules is one of the main 
achievements of this work. 
The construction of a full  event generator 
which uses this  approach,  
e.g. by 
extending the methods of~\cite{Baranov:2021uol} to the small-$x$ 
phase space~\cite{Monfared:2019uaj},   
will be the subject of future work. 
Such an event generator could be compared with 
existing small-$x$ Monte Carlo  
generators, e.g.~\cite{Chachamis:2015zzp,Andersen:2011zd,Jung:2010si,Hoeche:2007hlb,Golec-Biernat:2007tjf,Marchesini:1992jw,Andersson:1995jt,Orr:1997im,Schmidt:1996fg}.  
Also, fits of TMDs to experimental data 
based on the new evolution equation, using the xFitter 
platform~\cite{xFitterDevelopersTeam:2022koz,Alekhin:2014irh}, 
 are in progress.

\vskip 0.3 cm 

\noindent {\bf Acknowledgments}. The results presented 
have been obtained in collaboration with F.~Hautmann, M.~Hentschinski, 
A.~Kusina, K.~Kutak and A.~Lelek. I thank the conference organizers and 
convenors  for the invitation and the very interesting meeting.

\end{document}